\documentclass[a4paper]{article}

\usepackage[english]{babel}
\usepackage{authblk}
\usepackage[letterpaper,top=2cm,bottom=2cm,left=3cm,right=3cm,marginparwidth=1.75cm]{geometry}
\usepackage{amsmath}
\usepackage{amssymb}
\usepackage{graphicx}
\usepackage[colorlinks=true, allcolors=blue]{hyperref}
\usepackage{braket}
\usepackage{amsfonts}
\usepackage{qcircuit}
\usepackage[indent]{parskip}
\usepackage{algpseudocode}
\usepackage[misc]{ifsym}

\makeatletter

\title{Simulating NMR Spectra with a Quantum Computer}
\author[1]{J. Ossorio-Castillo \thanks{Email: joaquin@mestrelab.com (\Letter).}}
\author[1]{A. Rodríguez-Coello}
\affil[1]{\small{\textit{Mestrelab Research, Rúa de Feliciano Barrera Fernández, 9, 15706 Santiago de Compostela, Spain}}}

\begin{document}
\maketitle

\begin{abstract}
The procedure for simulating the nuclear magnetic resonance spectrum linked to the spin system of a molecule for a certain nucleus entails diagonalizing the associated Hamiltonian matrix. As the dimensions of said matrix grow exponentially with respect to the spin system's atom count, the calculation of the eigenvalues and eigenvectors marks the performance of the overall process. The aim of this paper is to provide a formalization of the complete procedure of the simulation of a spin system's NMR spectrum while also explaining how to diagonalize the Hamiltonian matrix with a quantum computer, thus enhancing the overall process's performance. Two well-known quantum algorithms for calculating the eigenvalues of a matrix are analyzed and put to the test in this context: quantum phase estimation and the variational quantum eigensolver. Additionally, we present simulated results for the later approach while also addressing the hypothetical noise found in a physical quantum computer.
\end{abstract}

\section{Introduction}

Accurate and efficient simulation of nuclear magnetic resonance spectra from a molecule's spin system requires constructing the Hamiltonian matrix of said spin system, whose dimensions increase exponentially with respect to the number of corresponding atoms in the molecule (see, for example, \cite{cobas2015integrated}). Subsequently, this Hamiltonian matrix is diagonalized to calculate its eigenvalues and eigenvectors, and the spectrum is simulated based on these values. It is precisely the diagonalization of the Hamiltonian matrix that poses a bottleneck in the current algorithm, as its performance scales exponentially with respect to the size of the molecule's spin system.

To mitigate the problems caused by this scaling, there are several existing techniques that permit us to calculate that spectrum when the dimensions are intractable (for example, fragmenting the spin system in several subsystems); however, the accuracy of the simulated spectrum is reduced as these techniques are applied. Therefore, an accurate and efficient alternative to this calculation would be of particular interest in the field of nuclear magnetic resonance spectroscopy.

The proposed project aims to achieve accurate and efficient simulation of an NMR spectrum using a quantum computer. Starting with known values for the chemical shifts and coupling constants of the spin system, an algorithm has been developed to calculate the free induction decay (FID) signal of the molecule in question, as well as its NMR spectrum (which is essentially the Fourier transform of the FID signal). More details on this can be found in \cite{keeler}.

Two main algorithms for tackling this task are proposed and analyzed: one based in quantum phase estimation \cite{kitaev1995quantum} and the other in the variational quantum eigensolver \cite{peruzzo2014variational}. The last one is especially relevant in the context of this project as, apart from laying the foundations of the simulation of NMR spectra using quantum computers, it also allows the possibility to test it with small molecules by means of the new quantum computer that has been recently installed at the Galicia Supercomputing Center (CESGA) in Santiago de Compostela, Spain.
\section{Constructing the Hamiltonian Matrix of a Spin System}\label{sec:constr-hamiltonian}

In this section, we introduce and explain the Hamiltonian matrix of a quantum spin system. A spin system is a collection of quantum particles possessing spin, undergoing interaction. Our proposed Hamiltonian is based in the quantum Heisenberg model for spin systems, with the particularity of taking into account all possible interactions between nuclei, and also the presence of individual terms due to the excitation of the spins in the presence of an external magnetic field.

\subsection{Brief Quantum Formalism}

There are two key ideas that one must consider when dealing with the quantum realm. In quantum mechanics, each possible state is represented by a state vector. Despite its abstract nature, a state vector is simply a mathematical representation containing all the information of the system. Consequently, each quantum system can be associated with a Hilbert space –-a vector space over the complex field--– whose elements are these state vectors.

On the other hand, physical properties are represented by Hermitian operators over the Hilbert space of the system. For example, let us consider a system consisting of a single particle with spin $s = 1/2$. The eigenvectors of its associated state space, i.e., its eigenstates, are $| \alpha \rangle$ and $| \beta \rangle$, often named the spin-up and spin-down states. Any other possible state will be a linear combination of these two states, obeying a normalization rule.

Now, this particle possesses a certain magnetic moment due to its spin. This quantity is associated with the spin operators:
\begin{equation*}
\vec{S} = (S_x, S_y, S_z).
\end{equation*}
The spin operator $S_z$ acts over the eigenvectors as:
\begin{equation*}
    S_z | \alpha \rangle = \frac{1}{2} | \alpha \rangle \hspace{5mm} S_z | \beta \rangle = -\frac{1}{2} | \beta \rangle 
\end{equation*}
Remarkably,  its eigenvalues are the two possible values of $m$ for a spin-1/2 particle, i.e. $m= -1/2, 1/2$.

Let us suppose now that the system is composed of two spin-1/2 particles. Each individual particle has an associated state space, namely $\mathcal{H}_1$ and $\mathcal{H}_2$. However, the total system also has its own state space. This global state space is the tensor product of the individual spaces: $\mathcal{H}_1 \otimes \mathcal{H}_2$. Thus, if both $\mathcal{H}_1$ and $\mathcal{H}_2$ were 2-dimensional, the global space would be 4-dimensional.

State vectors of this global space are simply pairs of state vectors of $\mathcal{H}_1$ and state vectors of $\mathcal{H}_2$ related via the Kronecker product. Thus, the eigenvectors will be: $\{ | \alpha \alpha \rangle , | \alpha \beta \rangle,  | \beta \alpha \rangle , | \beta \beta \rangle \}$. In this notation, $| \alpha \alpha \rangle$ means $| \alpha  \rangle \otimes | \alpha \rangle$.

Returning to the spin operator $S_z$ from the previous example, it is evident that an upgrade is necessary. Given the presence of two particles, there are now two spin operators, $S_{1z}$ and $S_{2z}$, both acting on states within the global $\mathcal{H}_1 \otimes \mathcal{H}_2$ and each providing information about an individual particle.

It is important to emphasize the following property of the Kronecker product: let $A$ and $B$ be two operators, and let $v$ and $w$ be vectors, the subsequent relationship is observed:
\begin{equation}
    \label{tensorproductproperty}
    (A \otimes B)(v \otimes w) = (Av) \otimes (Bw).
\end{equation} where $A \otimes B$ is the Kronecker product between matrices $A$ and $B$. This allows us to deduce the construction of the new operators $S_{1z}$ and $S_{2z}$ in the global space:
\begin{align*}
\begin{split}
S_{1z} = S_z \otimes I \vspace{5mm}
\\
S_{2z} = I \otimes S_z
\end{split}
\end{align*}

For a system of $k$ particles, considering that the overall system is represented by the tensor product of $k$ individual Hilbert spaces, the spin operator for an individual particle, $S_{iz}$, can be expressed as a Kronecker product of $k$ factors, where all factors are the identity matrix except for the one at position $i$, which corresponds to the actual spin operator $S_z$. We introduce the notation $(\mathbb{C}^2)^{\otimes k}$ to signify a tensor product of $k$ vector spaces. Thus, we can formulate $$S_{iz} = I ^{\otimes i-1} \otimes S_{z} \otimes I^{\otimes k-i}.$$This generalizes to any other operator acting on a subset of the overall system.

\subsubsection*{Note on spin operators}

A closer examination of the various spin operators can be provided. Note how  $\vec{S}$ is defined as a vector operator. Naturally, even in quantum physics, magnetic moment remains a vector quantity. The components of $\vec{S}$ encompass the three spin operators associated with each Cartesian axis. Notably, $S_z$ is conventionally chosen to align with the $Z$-axis, which corresponds to the direction of applied magnetic fields. The remaining two spin operators are defined such that the following commutation relationship (and cyclic permutations) holds: $$ [S_x, S_y] = i S_z.$$ For a spin-1/2 particle, spin operators can be expressed by means of the Pauli matrices:

\begin{equation*}
S_x = \dfrac{\hslash}{2} \sigma_x
\text{, }
S_y = \dfrac{\hslash}{2} \sigma_y
 \text{, and }
S_z = \dfrac{\hslash}{2} \sigma_z
\text{.}
\end{equation*}
where the Pauli matrices are defined as
\begin{equation*}
\sigma_x =
\begin{pmatrix}
0 & 1 \\
1 & 0
\end{pmatrix}\text{, }
\sigma_y =
\begin{pmatrix}
0 & -i \\
i & 0
\end{pmatrix} \text{, and }
\sigma_z =
\begin{pmatrix}
1 & 0 \\
0 & -1
\end{pmatrix}\text{.}
\end{equation*}
Additionally, the set of these matrices along with the identity matrix, as well as any combination via Kronecker product of them, form a group of operators. They are referred as the Pauli operators and are essential in quantum computing.

\subsubsection*{Hamiltonian operator}

Once the concept of quantum operator has been shown, we shall now introduce the Hamiltonian operator. The Hamiltonian of a system contains the key information of the system's energy and rules its evolution in time. Of course, in the quantum realm, this Hamiltonian is an operator whose eigenvalues correspond to the possible energy values of the system. As any other quantum operator, it may be associated with a matrix, which, in this case, is called the Hamiltonian matrix.

\subsection{Hamiltonian of a Spin System}\label{ssec:hamilt}

Let us consider a system of a few spin-1/2 nuclei under a magnetic field oriented along the $Z$-axis. As each particle possesses an intrinsic magnetic moment (referred to as its nuclear spin), the applied magnetic field will excite each individual particle. Consequently, when constructing the Hamiltonian matrix, this individual contribution must be carefully taken into consideration. This term is referred to as the Zeeman interaction. For an individual particle, the contribution will be $- \gamma (1 + \delta) B m$ \cite{keeler}. Breaking down this expression, we have $\gamma$ as the gyromagnetic ratio (which represents the relationship between its intrinsic angular momentum and its intrinsic magnetic moment), $\delta$ as the chemical shift (a parameter related to the electronic structure of a particle due to its chemical environment), $B$ as the magnitude of the applied field, and finally, the quantum number $m$. Overall, this can be viewed as simply as the product between the magnetic moment of the particle and the applied field. Keeping in mind that we are only interested in the $Z$-axis, we can express the aforementioned expression as $w S_z$, where $w$ encompasses all the scalar components. Thus, summing up all the individual contributions from the $k$ nuclei, this addend of the Hamiltonian is expressed as follows:
\begin{equation*}
    H_Z = \sum _k w_k S_{kz}.
\end{equation*}

Another significant energetic contribution is present: nuclei can be coupled, implying that the spin of each nucleus interacts with the spins of the surrounding nuclei. This interaction becomes easier to understand if we envision each nucleus as a small magnet that interacts with other magnets. Within this framework, we can easily calculate the energy of two spins, or two dipoles. Similar to the contribution of the Zeeman effect, the energy will be a product of the magnetic moment of particle 1 and the magnetic field of particle 2. Unlike the presented Zeeman case, we cannot assume the magnetic field to be oriented along the $Z$-axis. In fact, this magnetic field obeys a function dependent on the position of particle 2 and its intrinsic magnetic moment. Also, it is important to note that now the term magnetic moment refers to the complete vector magnitude. When we evaluate the product of both the magnetic moment and the magnetic field of a spin particle, we arrive at the following expression:
$$E = J \vec{m_1} \vec{m_2}.$$ In other words, the energy contribution is proportional to the product of both magnetic moments, being $J$ the interaction amplitude.

It should already be clear that $\vec{m_1}$ and $\vec{m_2}$ correspond to $\vec{S_1}$ and $\vec{S_2}$, respectively. On the other hand, these operators are connected through the tensor product. Remember that the energy contribution is the product of the magnetic moment of particle 1 and the magnetic moment of particle 2. The involvement of the tensor product in operators that describe properties of individual particles within a global system has already been discussed. Thus, as stated in \cite{blundell2014magnetism}, the Hamiltonian term of two coupled spins can be written  as:
\begin{equation*}
    H_J = J \vec{S_1} \cdot \vec{S_2}.
\end{equation*}

If we take into account anisotropy, the magnitude $J$ should become a tensor that provides details about how the amplitude differs in each direction. However, under the assumption of isotropy, $J$ becomes a scalar. Additionally, we can clarify that the scalar product of the vector operators refers to the Kronecker product of their components. Thus, we can rewrite the expression as follows:
\begin{equation*}
    H_J = J (S_{1x} \otimes S_{2x} + S_{1y} \otimes S_{2y} + S_{1z} \otimes S_{2z} )
\end{equation*}

This coupling interaction needs to be extended to the case of multiple nuclei. A question that might arise is how many neighboring nuclei should be taken into consideration. However, this matter will not be discussed here. In our case, all interactions will be computed.

Consider a system with three coupled spins. Unlike the scenario of a two-particle coupling, there are now three distinct couplings: nucleus 1 interacts with nuclei 2 and 3, while nucleus 2 and nucleus 3 also interact between themselves. In this context, a single $J$-value no longer applies. Instead, each coupling possesses its own unique $J$-value, denoted as $J_{ij}$ when nuclei $i$ and $j$ are involved. Observe that the product of operators becomes somewhat more intricate, yet it remains straightforward to articulate. With three different couplings, there arise three distinct products. Let us explicitly express the contribution of each coupling to the Hamiltonian:
\begin{align*}
\begin{split}
    H_{J12} = J_{12}(\vec{S_1} \otimes \vec{S_2} \otimes I)
    \\
    H_{J13} = J_{13}(\vec{S_1} \otimes I \otimes \vec{S_3})
    \\
   H_{J23} = J_{23}(I \otimes \vec{S_2} \otimes \vec{S_3})
\end{split}
\end{align*}

Once again, observe how these operators are constructed through tensor products of the identity matrix $I$ and the spin operators, positioned according to the nuclei indices. Thus, for a spin system comprising $k$ nuclei, the coupling term can be expressed as:
\begin{equation*}
    H_J = \sum _{i,j} J_{ij} \vec{S_i} \cdot \vec{S_j}
\end{equation*} It is important to note that the sum runs twice for each coupling pair. However, the definition of the $J$-value holds some ambiguity and may incorporate a factor of $1/2$.

Finally, we arrive at the expression for the Hamiltonian of a spin system, considering the couplings among all spins:
\begin{equation*}
     H = \sum _k w_k S_{kz} + \sum ^{k} _{i,j} J_{ij} \vec{S_i} \cdot \vec{S_j}
\end{equation*}
It is worth mentioning that the dimensions of the Hamiltonian matrix are $2^k \times 2^k$, corresponding to the $2^k$-dimensional Hilbert space of the entire system. For a possible implementation of the process of calculating the Hamiltonian matrix of a spin system in Python we refer to the code written by our colleague Gary Sharman, which can be found in \cite{sharman_gary_2023_8337288}.
\section{Quantum Algorithm for Calculating the Eigenvalues of a Hamiltonian Matrix}

Now that we have a way to construct our Hamiltonian matrix directly from the chemical shifts and the $J$-couplings of our quantum system for a certain magnetic field, it is time to analyze the possibilities of diagonalizing such matrix, thus obtaining the eigenvalues and eigenvectors of the system. We have analyzed two main alternatives from the current state of the art: the quantum phase estimation algorithm, and the variational quantum eigensolver. In the present chapter we will describe both approaches, while also enumerating all the advantages and setbacks found for each of them.

\subsection{Approach 1: Quantum Phase Estimation}\label{ssec:qpe}

Quantum phase Estimation (or QPE for short) was originally described in \cite{abrams1999quantum} and can be formulated as follows. Given a unitary operator $U$ acting on $m$ qubits (i.e., a unitary matrix of dimension $2^m \times 2^m$), then all its eigenvalues have unit modulus. Thus, if $\ket{\varphi}$ is an eigenvector of $U$, then $U \ket{\varphi} = e^{2 \pi i \theta} \ket{\varphi}$ for some $\theta \in \mathbb{R}$ (we can assume that $0 \leq \theta < 1$). The idea of the algorithm is to obtain $\theta$ with high probability, thus obtaining the eigenvalue associated with the eigenvector $\ket{\varphi}$.

The number of quantum logic operations required to estimate the eigenvalue with precision $\epsilon$ is $\mathcal{O}(1/\epsilon)$, while the number of additional qubits (apart from the ones used to describe $\ket{\varphi}$) is $\mathcal{O}(\log(1/\epsilon))$ \cite{abrams1999quantum}. Although as pointed out in \cite{peruzzo2014variational} this approach requires fully coherent evolution, which has not been reached yet in current quantum computers, we will nevertheless analyze the possibilities of applying it to our problem.

\paragraph{$\mathbb{SETUP}$}
$$$$\noindent\framebox{\parbox[b]{\linewidth}{\begin{algorithmic}
\State $\Ket{\psi_0}_{t,m} \leftarrow \Ket{0}_t \Ket{\varphi}_m$
\end{algorithmic}}} \\

If $\ket{\varphi}$ is an eigenvector of $U$ and $e^{2  \pi i \theta}$ is its associated eigenvalue, the QPE algorithm is capable of estimating the phase $\theta$. In order to do that, the algorithm is implemented via two different quantum registers. The first one contains $t$ ancilla qubits which may be referred as the estimation qubits, whereas the second register is commended to prepare the state $\ket{\varphi}$ and perform the controlled-$U$ operation using $m$ qubits.  Alternatively, we may see both registers as the control and target ones.

\paragraph{$\mathbb{STEP}$ 1}
$$$$\noindent\framebox{\parbox[b]{\linewidth}{\begin{algorithmic}
\State $\Ket{\psi_1}_{t,m} \leftarrow \left( \boldsymbol{H}^{\otimes t} \otimes \boldsymbol{I}^{\otimes m} \right) \ket{\psi_0}_{t,m} $
\end{algorithmic}}} \\

Once the initial state has been prepared, the first step consists in applying Hadamard gates to the control qubits, obtaining the following global state:
\begin{equation*}
    \ket{\psi_1}_{t,m} = \frac{1}{\sqrt{2^t}} \sum _{k=0}^{2^t -1} \ket{k}_t \ket{\varphi}_m
\end{equation*}
We may clarify that $1/\sqrt{2^t}$ arise as a normalization fraction and that $\ket{k}$ corresponds to the global state of the first register (as each ancilla qubit is either in state $\ket{0}$ or $\ket{1}$, we can understand the target register as the integer $k$ equivalent to the binary string formed by each qubit value).

\paragraph{$\mathbb{STEP}$ 2}
$$$$\noindent\framebox{\parbox[b]{\linewidth}{\begin{algorithmic}
\State $\Ket{\psi_2}_{t,m} \leftarrow  \boldsymbol{C}_{U} \ket{\psi_1}_{t,m} $
\end{algorithmic}}} \\

Now the system undergoes a process of controlled-$U$ gates being applied to the second register. Assigning to each control qubit an index $j$, the $j$ qubit controls the application of the $U$ gate $2^{j}$ times. Since $U^{2^j} \ket{\varphi} = e^{2 \pi i 2^j \theta} \ket{\varphi}$, via the phase kick-back phenomenon the state of both registers after this process is: $$\ket{\psi_2}_{t,m} = \frac{1}{\sqrt{2^t}} \sum _{k=0}^{2^t-1} e^{2 \pi i k \theta} \ket{k}_t \ket{\varphi}_m$$

From now on, we can forget about the second register and just take into account the $t$ qubits of the first register. $$\ket{\psi_2}_{t} = \frac{1}{\sqrt{2^t}} \sum _{k=0}^{2^t-1} e^{2 \pi i k \theta} \ket{k}_t $$

\paragraph{$\mathbb{STEP}$ 3}
$$$$\noindent\framebox{\parbox[b]{\linewidth}{\begin{algorithmic}
\State $\Ket{\psi_3}_{t} \leftarrow  \boldsymbol{QFT}^{-1}_t \ket{\psi_2}_{t} $
\end{algorithmic}}} \\

Performing the inverse quantum Fourier transformation to the first register of qubits, the state transforms to:
\begin{equation*}
\begin{split}
\Ket{\psi_3}_{t} & = \frac{1}{\sqrt{2^t}} \sum _{k=0}^{2^t-1} e^{2 \pi i k \theta} \left( \frac{1}{\sqrt{2^t}}\sum_{x=0}^{2^t - 1} e^{\frac{-2\pi i kx}{2^t}}\ket{x}_t \right) \\
 & = \frac{1}{2^{t}}\sum_{x=0}^{2^t - 1} \sum_{k=0}^{2^t - 1}e^{-\frac{2\pi i k}{2^t} \left ( x - 2^t \theta \right )}  \ket{x}_t
\end{split}
\end{equation*}

For the sake of simplicity, we can rewrite that as$$\Ket{\psi_3}_{t} = \sum_{x=0}^{2^t - 1} \alpha_x \ket{x}_t $$ where $$\alpha_x = \frac{1}{2^{t}} \sum_{k=0}^{2^t - 1}e^{-\frac{2\pi i k}{2^t} \left ( x - 2^t \theta \right )}.$$
If we round $2^t \theta$ to its nearest integer (namely $a \in \mathbb{Z}$), then $2^t \theta = a + 2^t \delta$ where $0 \leq \left| 2^t \delta \right| \leq 1/2$. Thus: $$\alpha_x = \frac{1}{2^{t}} \sum_{k=0}^{2^t - 1} e^{-\frac{2\pi i k}{2^t} \left ( x - a \right )} e^{2 \pi i \delta k}.$$

\paragraph{$\mathbb{STEP}$ 4}
$$$$\noindent\framebox{\parbox[b]{\linewidth}{\begin{algorithmic}
\State $\tilde{x} \leftarrow \text{measure the first register}$
\end{algorithmic}}} \\

We are finally in position to measure the $t$ qubits of the first register, obtaining an integer $\tilde{x}$ between 0 and $2^t - 1$. If we analyze the probability of retrieving a certain $\tilde{x}$, we obtain:
$$P(\tilde{x}) = \left| \alpha_x \right|^2 = \left| \frac{1}{2^{t}} \sum_{k=0}^{2^t-1} e^{\frac{-2\pi i k}{2^t}(x-a)} e^{2 \pi i \delta k} \right |^2.$$

Notably, if $2^t \theta$ is already an integer, then $\delta = 0$ and $P(a) = 1$, so we get directly the exact value of $\theta$ we are looking for on the first try. However, the general case, in which $2^t \theta$ is not an integer, needs additional insight. Let us write the probability of obtaining the desired number in that case:
$$P(a) = \frac{1}{2^{2t}} \left| \sum_{k=0}^{2^t-1} e^{2 \pi i \delta k} \right |^2 \geq \dfrac{4}{\pi^2} \approx 0.4052847...$$
The proof of the last inequality can be found in \cite{cleve1998quantum}, which gives an acceptable probability of success. Moreover, as stated in \cite{Nielsen2010quantum}, it is possible to obtain $\theta$ with and accuracy of $2^{-n}$ and with a probability of success of at least $1-\varepsilon$ just by forcing the number of estimation qubits $t$ to be
\begin{equation*}
    t = n + \left\lceil \log({2+\frac{1}{2\varepsilon})} \right\rceil.
\end{equation*}

Now, it is obvious that the condition of preparing the second register with the state $\ket{\varphi}$ requires to already know said eigenvector. A state $\ket{V}$ may be prepared such that $|\braket{V|\varphi} |^2$ relates to the probability of the QPE still having success. More precisely, $\ket{V}$ may be expressed as a linear combination of the actual eigenvectors of the operator $U$., i.e., $\ket{V} = \sum _m c_m \ket{\varphi_m}$. In that case, the output of the algorithm will be one of the $m$ possible eigenvalues, and the probability of extracting each one is $|c_m|^2$. 

Thus, by preparing initial $\ket{V}$ states which satisfy $|\braket{V|\varphi} |^2 \approx$ 1, and performing multiple measurements, QPE is able to obtain all eigenvalues $\varphi_m$, each one with a probability of success of $|c_m|^2(1-\varepsilon)$.

Focusing now on our problem, given the Hamiltonian matrix $H$, we may build the unitary operator $U = e^{i2\pi H}$. It is well known that $H$ and $U$ share eigenvectors. Moreover, their eigenvalues are related via complex exponentiation. If $\lambda$ is an eigenvalue of $H$, then $e^{i2\pi \lambda}$ is an eigenvalue of $U$. Thus, by retrieving the phases of the eigenvalues of $U$, we obtain the eigenvalues of $H$.

In order to build the QPE circuit, a quantum gate for $U$ must be provided. Thanks to the Hamiltonian being naturally encoded as Pauli operators, this task is fairly achievable.

First, we may decompose our global Hamiltonian as a sum of terms acting on a reduced number of qubits. Although, on a strictly mathematical sense, addends and sum all have identical dimension, we may take advantage of how each addend only refers to a two body interaction at most. Furthermore, since all terms are Hermitian we may write:

\begin{equation*}
    e^{it\sum H} = \lim\limits_{r \to \infty} (e^{itH_1}e^{itH_2}...e^{itH_{n-1}}e^{itH_n})^r.
\end{equation*}

This expression is known as the Lie-Trotter product formula \cite{trotter1959product}. Unless the matrices commute, the exponential of the sum of two matrices is not equal to the product of the individual exponentials. Thus, we must establish a limit, and the equality holds when $r$ tends to infinity ($r$ is known as the Trotter number, indicating the number of Trotter steps). Intuitively, Trotterization may be seen as slicing a big temporal process into many smaller processes, which by iteration lead to the same final result.

Since the product formula must be truncated by the choice of a finite $r$ value, we may investigate the error associated to the approximation. Given an approximation $U_1$ of $e^{it\sum_i H_i}$ with $r$ Trotter steps, the error is upper bounded by the following equation, where $|| \bullet ||$ denotes the Frobenius norm \cite{layden2022trottererror}:
\begin{equation*}
    ||e^{it \sum_i H_i}  - U_1|| \leq \frac{t^2}{2r}\sum _ {j=1}^{L} ||\sum _ {k=j+1}^{L} [H_k, H_j] ||.
    \label{trotterbound}
\end{equation*}

Now it is time to begin the translation of exponential operators into quantum gates. Firstly, we recall exponentiation of Pauli matrices are rotations on the Bloch sphere about $\vec{x}$, $\vec{y}$, $\vec{z}$. Therefore, the rotation quantum gates $R_x$, $R_y$, $R_z$ are defined as follows: 
\begin{equation*}
\begin{aligned}
    R_x(\theta) \equiv e^{-i\theta \sigma_x /2} = \cos{\frac{\theta}{2}I} - \sin{\frac{\theta}{2}\sigma_x} = \begin{bmatrix}
        \cos{\frac{\theta}{2}} & -i\sin{\frac{\theta}{2}} \\
        -i\sin{\frac{\theta}{2}} & \cos{\frac{\theta}{2}}
    \end{bmatrix} \\
    R_y(\theta) \equiv e^{-i\theta \sigma_y /2} = \cos{\frac{\theta}{2}I} - \sin{\frac{\theta}{2}\sigma_y} = \begin{bmatrix}
        \cos{\frac{\theta}{2}} & -\sin{\frac{\theta}{2}} \\
        \sin{\frac{\theta}{2}} & \cos{\frac{\theta}{2}}
    \end{bmatrix}\\
    R_z(\theta) \equiv e^{-i\theta \sigma_z /2} = \cos{\frac{\theta}{2}I} - \sin{\frac{\theta}{2}\sigma_z = \begin{bmatrix}
        e^{-i\theta /2} & 0 \\
        0 & e^{-i\theta/2}
    \end{bmatrix}}
\end{aligned}
\end{equation*}

In our case, we deal with the exponentiation of Kronecker products of Pauli matrices.  
We will make use of the following relationship for any  real number $t$ and any matrix $A$ such  that $A^2 = I$.
\begin{equation*}
    e^{iAt} = \cos{(t)}I + i\sin{(t)}A
    \label{exponentialofmatrixastrigo}
\end{equation*}
We note that for any tensor product of Pauli operators, namely $P = (\sigma_i \otimes ... \otimes\sigma_j ) $, it holds that $P^2 = I$.  The total tensor product can be squared making use of the property portrayed in Equation \ref{tensorproductproperty}. Thus, $P^2 = (\sigma_i \otimes ... \otimes\sigma_j ) \cdot (\sigma_i \otimes ... \otimes\sigma_j ) = (\sigma_i \sigma_i \otimes ... \otimes\sigma_j \sigma_j )$. Now $\sigma ^2 = I$ for any $\sigma \in \{I, \sigma_x, \sigma_y, \sigma_z \} $, and trivially the Kronecker product of identity matrices is equal to the identity matrix of the product space.

Now we are able to write $e^{i(I \otimes Z)t} = I  \otimes R_z(-2t)$ and, analogously, $e^{i(Z \otimes I)t} = R_z(-2t) \otimes  I$. Thus, a quantum circuit implementing the exponential of the individual terms of the Hamiltonian is achieved by simply applying a $R_z$ gate to the corresponding qubit.

In the case of having $e^{i(Z \otimes Z)t}$, the quantum circuit needs to be complemented with CNOT gates both before and after the $R_z$ gate. In other words, $e^{i(Z \otimes Z)t} = CNOT (I  \otimes R_z(-2t)) CNOT$, and the quantum circuit is portrayed as:

\[ \Qcircuit @C= 1em @R= .7em {
& \ctrl{1} & \qw & \ctrl{1} & \qw \\
& \targ & \gate{R_z} & \targ & \qw
} \]

The first CNOT entangles the two qubits. With this mental framework it is easy to generalize the former quantum circuit implementation for any $e^{i(... Z \otimes ... \otimes Z \otimes ...)}$ by modifying the indices of the qubits on which the gates act. Examples for $e^{i(Z \otimes I \otimes Z )}$ and $e^{i(I \otimes Z \otimes Z )}$ are given.

\begin{figure}[h!]
\centering
\sbox0{
\Qcircuit @C = 1em @R = .7em {
& \ctrl{2} & \qw & \ctrl{2} & \qw \\
& \qw & \qw & \qw & \qw \\
& \targ & \gate{R_z} & \targ & \qw
}
}
\sbox1{
 \Qcircuit @C = 1em @R = .7em {
& \qw & \qw & \qw & \qw \\
& \ctrl{1} & \qw & \ctrl{1} & \qw \\
& \targ & \gate{R_z} & \targ & \qw
} 
}
\begin{tabular}{c|c}
    \usebox0 & \usebox1
\end{tabular}
\end{figure}

Lastly, coupling terms involving operators $X$ and $Y$ may be implemented with the same model but performing a change of basis. This is achieved by means of the Hadamard gate to alternate between $X$ basis and $Z$ basis, and $Y \equiv R_x(-\pi/2)$ to alternate between $Y$ basis and $Z$ basis \cite{whitfield2011simulationelectronichamiltonian}.

Thus, we present the quantum circuit for $e^{i(X \otimes X)t}$
\[ \Qcircuit @C= 1em @R= .7em {
& \gate{H} & \qw & \ctrl{1} & \qw & \ctrl{1} & \qw & \gate{H} & \qw\\
& \gate{H} & \qw & \targ & \gate{R_z} & \targ & \qw & \gate{H} & \qw
} \]
and for $e^{i(Y \otimes Y)t}$
\[ \Qcircuit @C= 1em @R= .7em {
& \gate{Y} & \qw & \ctrl{1} & \qw & \ctrl{1} & \qw & \gate{Y^{\dag}} & \qw\\
& \gate{Y } & \qw & \targ & \gate{R_z} & \targ & \qw & \gate{Y^{\dag}} & \qw
} \]

With this procedure we can map the operator $e^{(i2 \pi H)}$ into a set of quantum gates that shall be bundled into the unitary gate $U$ composing the quantum phase estimation circuit. We may finally clarify the angle of rotation of the $R_z$ gates would be $-4\pi\theta$, with $\theta$ being the chemical shift or the $J$-value that accompanies the exponentiated operator.

Another aspect to take into account is that, since a value is indistinguishable from itself plus any number of $2 \pi$ radians rotations, QPE can only estimate phases within the interval [0,1). This means large eigenvalues are not tractable. As proposed in \cite{gomez2022matrixdefiniteness}, we can adjust the range of the eigenvalues of matrix $H$ by dividing said matrix by a constant $C$ dependent on the bounds of the maximum and minimum eigenvalues.  The estimation of these bounds requires the trace of $H$ and the trace of $H^2$. However it can be shown that there is no need to build the whole matrices to compute these parameters, and instead both traces can be calculated making use of the Pauli operators and the mathematical relationships $\text{tr}(A \otimes B) = \text{tr}(A)\text{tr}(B)$ and $\text{tr}(A+B) = \text{tr}(A) + \text{tr}(B)$.

In our case, we chose $C$ to be four times the bound of the greater eigenvalue. This way, the eigenvalues will be within the range [-0.25, 0.25]. However, negative eigenvalues still pose a problem for the QPE algorithm. It is then desirable to shift the range bounds towards [0, 0.5], which can be achieved by adding an extra phase contribution to each ancilla register equal to $0.25$ times the number of $U$ gates the corresponding ancilla qubit controls. We may note that this scaling is translated into the argument of the rotation gates undergoing the same scale transformation. Once the algorithm estimates a phase, it is corrected by subtracting the additional $0.25$ and later multiplying by $C$ to obtain the desired eigenvalue.

With the QPE algorithm explained we may propose the following workflow. We run the QPE algorithm a certain number of trials for a random initial state until all eigenvalues have been obtained (it must be noted that QPE does not necessarily return an actual eigenvalue as, even if the algorithm is built to maximize the odds of returning an eigenvalue, other outcomes are still possible). Thus one should have a efficient method to perform this verification after each step. 
\subsection{Approach 2: Variational Quantum Eigensolver}\label{ssec:vqe}

The variational quantum eigensolver ---from now on, VQE--- was first introduced in \cite{peruzzo2014variational} as an alternative to quantum phase estimation, and requires combining a quantum processor with a classical computer. Its main advantage is that the requirements of such quantum processor can already be found in the current state of quantum computing, usually referred to as the noisy intermediate-scale quantum ---NISQ--- era (for example, the depth of the circuit does not increase as in with quantum phase estimation, and consequently the noise does not affect the algorithm in the same way). The VQE  belongs to a family of algorithms called VQAs (short for variational quantum algorithms), which make use of classical optimization to optimize the parameters of the quantum circuit involved in the process.

As explained in \cite{cerezo2021variational}, the first step of any variational quantum algorithm consists in defining a cost function that encodes the solution of the problem. In the case of VQE, it has the following shape: $$C(\boldsymbol{\theta}) = \bra{\psi (\boldsymbol{\theta})} H \ket{\psi (\boldsymbol{\theta})}.$$ This cost function fulfills the inequation $E_G \leq C(\boldsymbol{\theta})$, where $E_G$ is the energy of the ground state of the system (i.e., the lowest eigenvalue of the Hamiltonian matrix). Thus, whenever $\ket{\psi (\boldsymbol{\theta}_i)}$ equals the ground state of $H$, namely $\ket{\psi_G}$, for a certain $\boldsymbol{\theta}_i$, it holds that $C(\boldsymbol{\theta}_i) = E_G$.

The idea now is to write the Hamiltonian as a linear combination of products of Pauli operators ---which we already did in Section \ref{sec:constr-hamiltonian}--- so that it becomes measurable by a quantum computer. Let us suppose that such representation of our Hamiltonian matrix is $$H = \sum_k c_k P_k, $$where $P_k \in \{I, \sigma_x, \sigma_y, \sigma_z\}^{\otimes n}$ are the aforementioned products of Pauli operators (for $n$ being the number of qubits of our quantum system) and $c_k \in \mathbb{R}$ are a set of linear coefficients. Then, the minimization problem to be resolved can be stated as: $$\min_{\boldsymbol{\theta}} \sum_k c_k \bra{0}_n U^{\dag}(\boldsymbol{\theta}) \,P_k \, U(\boldsymbol{\theta}) \ket{0}_n$$

Finally, if we define $E_{P_k}(\boldsymbol{\theta}) = \bra{0}_n U^{\dag}(\boldsymbol{\theta}) \,P_k \, U(\boldsymbol{\theta}) \ket{0}_n$, we can see that $E_{P_k}$ is the expected value of the term $P_k$, which can be computed by a quantum processor. If the calculation of $E_{P_k}(\boldsymbol{\theta})$ is seen as a black box, then we obtain a classical minimization problem whose objective function is $$\min_{\boldsymbol{\theta}} \sum_k c_k E_{P_k}(\boldsymbol{\theta}).$$

Then, we need to propose an ansatz $U(\boldsymbol{\theta})$ ---i.e., the structure for the parameterized quantum circuit---, an initial state $\ket{\psi_0}$ and a trial state $\ket{\psi (\boldsymbol{\theta})} = U(\boldsymbol{\theta}) \ket{\psi_0}$. Given those, the expected value of the Hamiltonian can be calculated as:

\begin{equation*}
\begin{split}
E(\boldsymbol{\theta}) & = \bra{\psi(\boldsymbol{\theta})}\, H \, \ket{\psi(\boldsymbol{\theta})} \\
 & = \bra{\psi_0} \, U^{\dag}(\boldsymbol{\theta}) \, H \, U(\boldsymbol{\theta}) \, \ket{\psi_0} \\
 & = \sum_k c_k \bra{\psi_0} \, U^{\dag}(\boldsymbol{\theta}) \, P_k \, U(\boldsymbol{\theta}) \, \ket{\psi_0}
\end{split}
\end{equation*}

Although the most common ansatz in the literature is the Unitary Coupled Cluster (UCC) ansatz (see, for example: \cite{omalley2016scalable}) it is not directly applicable to our case because of the nature of the Hamiltonian, as it requires to have said Hamiltonian expressed in terms of creation and annihilation operators, while ours is expressed in terms of Pauli operators. There may exist the possibility of applying a Jordan-Wigner transformation to our Hamiltonian to overcome this hurdle \cite{nielsen2005fermionic}, but we have not explored that path. Instead, we have followed \cite{jattana2022assessment} and \cite{jattana2023improved} for constructing a different ansatz.

For a system of $N$ spins our proposed ansatz, the XY-ansatz, is:

\begin{equation*}
    U(\boldsymbol{\theta}) = [\prod _{l=N-1}^{1} \prod _{k=N}^{l+1} U_{lk}(\theta _{lk})] [\prod _{l=N-1}^{1} \prod _{k=N}^{l+1} U_{kl}(\theta _{kl})] 
\end{equation*}
where:
\begin{equation*}
    U_{pq}(\theta _{pq}) = \left\{ 
    \begin{array}{lc} e^{-i\theta _{pq} \sigma_{p}^y\sigma_q^x} & \text{if $p=N$ or $q=N$}\\
    e^{-i\theta _{pq} \sigma_{p}^y\sigma_q^x \sigma_N^z} & \text{otherwise}
    \end{array} \right.
\end{equation*}
We clarify $\sigma ^i _p$ refers to the $i$ Pauli operator acting on the $p$ qubit. For the said system of $N$ spins, the ansatz will consist of $N(N-1)$ parameters to be optimized.
 
Now in order to compute the expected value of a certain Pauli operator, a customized quantum circuit must be constructed consisting of the ansatz part and a measurement part. Notice that all measurements are done in the computational basis, and consequently some changes of basis may be required. For the $X$ operator a Hadamard gate must be applied whereas for a the $Y$ operator the $S^{\dag}$ is needed. Computation of expected values requires the quantum circuit to be run several times. In order to achieve a result with precision $\epsilon$, $\mathcal{O}(1/{\epsilon^2})$ shots are needed. The total number of quantum circuit executions may be reduced by measuring more than one operator's expected value at once. This can be performed as long as some commuting relationship is held \cite{tilly2022bestpractices}. 

Qubit-wise commuting refers to the commutation relationship held by Pauli operators of two different Hamiltonian terms acting on the same qubit space. For example, taking the terms $ZI$ and $ZZ$, both consist on the same Pauli operator acting on the firs qubit. As for the second qubit, $I$ and $Z$ commute, therefore only a quantum circuit is needed to measure both expected values. Another example would be terms such as $XIXI$ and $IXIX$ (or any other permutation) when we may compute several expected values with just one quantum circuit by simply selecting which qubits are measured. Of course, global commuting is still a valid criterion for grouping terms. This procedure is known as Pauli grouping; however, it should be noted that finding the optimal Pauli grouping is far from trivial.

Since in order to diagonalize the Hamiltonian matrix all eigenvalues and eigenvectors must be known, the VQE algorithm must be improved somehow. The folded  spectrum method proposed in \cite{tazi2023folded} is capable of finding all eigenvalues by means of computing the expected value of $(H-wI)^2$.  The key idea is that $H$ and $(H-wI)^2$ share eigenvectors but have their eigenvalues in different order. The ground state of $(H-wI)^2$ coincides with one of the excited states of $H$, specifically with the one with eigenvalue closest to $w$. 

That being said, the cost function shall be:
\begin{equation*}
    E(\boldsymbol{\theta}) = \bra{\psi(\boldsymbol{\theta})}\, (H-wI)^2 \, \ket{\psi(\boldsymbol{\theta})}.
\end{equation*}
And the VQE must be performed for a range of $w$ values. However, this method presents the drawback of increased number of operators to be measured, and thus Pauli grouping is desired in order to manage this issue.

Another approach consists on modifying the cost function by adding a new term  related to the condition that each new eigenstate must be orthogonal to the previous ones \cite{higgott2019vqe_all_states}.  For the $k$-th state, the cost function would be: 
\begin{equation*}
    E(\boldsymbol{\theta}_k)  = \bra{\psi(\boldsymbol{\theta}_k)}\, H \, \ket{\psi(\boldsymbol{\theta}_k)} + \sum _{i=0}^{k-1} \beta _i \left|\langle \psi(\boldsymbol{\theta}_k)| \psi(\boldsymbol{\theta}_i) \rangle \right|^2
\end{equation*}
This new term is zero if the states are orthogonal. The coefficient $\beta _i$ needs to be large enough, fulfilling the condition $\beta _i > \lambda _k - \lambda _i$.  This condition arises from the following reasoning: suppose we want to find the $k$-th eigenvalue of the Hamiltonian H. This problem is equivalent to finding the ground state of
\begin{equation*}
    H_k = H + \sum _{i=0}^{k-1} \beta _i \ket{i} \bra{i},
\end{equation*}
with $\ket{i}$ being the previous eigenstates. Then, for an arbitrary $\ket{\psi}=\sum a_i \ket{i}$, we have
\begin{equation*}
    \bra{\psi} H_k \ket{\psi} = \sum _{i=0}^{k-1} |a_i|^2(\lambda_i + \beta_i) + \sum _{i=k}^{d-1}|a_i|^2\lambda_i,
\end{equation*}
where $d$ is the total number of eigenvectors of $H$. Taking wrongly $\beta_i = \gamma - \lambda _i \leq \lambda _k - \lambda _i$ , the minimization of the cost function will return $\gamma$ instead of the actual eigenvalue.

A quantum circuit to compute the value  $|\langle \psi(\boldsymbol{\theta}_k)| \psi(\boldsymbol{\theta}_i) \rangle|^2$  is proposed in \cite{circuitforoverlap}. Provided the gate composition of the ansatz, $A$, and its inverse, $A^{\dagger}$,  we may build a circuit applying $A(\boldsymbol{\theta}_k)^{\dagger}A(\boldsymbol{\theta}_i)$ to the state $\ket{0}$. The overlap is then estimated by counting the relative frequency of measuring again the state $\ket{0}$.

Within this method the workflow consists in computing the lowest eigenvalue and its eigenvector in the firs step as in a classic VQE, and then computing the next ones. For $\beta_i$ we may propose to choose the difference between the lower and upper bound of the eigenvalues.  In every computation not only the inherent errors of the VQE apply but also a phenomenon of error accumulation arises.

All these possible VQE implementations and variations share the classical process of committing an optimization, a step that presents its own issues. Remarkably, the cost function often presents local minima \cite{tilly2022bestpractices} which may lead to an outcome which is not a real eigenvalue. On the other hand, while minimizing the function, the gradients of the cost function may vanish exponentially with respect to the system size due to, for example, a random initialization of the circuit parameters, a phenomenon known as the barren plateau problem \cite{tilly2022bestpractices}. These issues must be taken in mind when selecting the optimization method: for example, gradient-free methods such as Simultaneous Perturbation Stochastic Approximation (SPSA) or COBYLA may be preferred \cite{failde2023using}. However, even gradient-free optimizers may be affected by the barren plateau problem \cite{Arrasmith2021effectofbarren}.

\section{Example: Sulfanol}

\begin{figure}[t]
    \centering
    \includegraphics[scale=0.3]{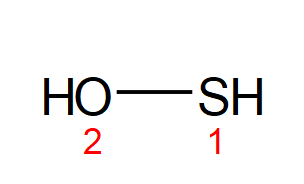}
    \caption{Sulfanol molecule}
    \label{fig:odcb-sulfanol}
\end{figure}

We proceed to illustrate the construction of a Hamiltonian matrix using the example of hydrogen thioperoxide ---also called sulfanol---, whose molecule, as rendered in MestReNova, is shown in Figure \ref{fig:odcb-sulfanol}. As can be seen, this molecule has the structure H–S–O–H, with two hydrogen atoms, one sulphur atom, and one oxygen atom. If we want to calculate the $^1$H-NMR spectrum of said molecule, we must take into account only the hydrogen atoms, thus obtaining a spin system of two particles.

After predicting the spin system --assuming a spectrometer frequency of 400 MHz-- and labeling the atoms as in Figure \ref{fig:odcb-sulfanol}, we obtain the following values for the chemical shifts and $J$-couplings:

\begin{equation*}
\begin{split}
\delta_1 & = 3.44 \text{ ppm} \\
\delta_2 & = 7.40 \text{ ppm} \\
J_{1,2} & = 2.32 \text{ Hz} \\
\end{split}
\end{equation*}

As explained in Section \ref{ssec:hamilt}, and after the agreement that $\hslash \approx 1$, the contributions of the different chemical shifts to the Hamiltonian matrix can be calculated as follows: 
\begin{equation*}
H_{\delta_1} = w_{\delta_1} \left( S_z \otimes I_2 \right) =  w_{\delta_1}
\begin{pmatrix}
1/2 & 0 \\
0 & -1/2
\end{pmatrix}
\otimes
\begin{pmatrix}
1 & 0 \\
0 & 1
\end{pmatrix}
\end{equation*} 
and
\begin{equation*}
H_{\delta_2} = w_{\delta_2} \left( I_2 \otimes S_z \right) =  w_{\delta_2}
\begin{pmatrix}
1 & 0 \\
0 & 1
\end{pmatrix}
\otimes
\begin{pmatrix}
1/2 & 0 \\
0 & -1/2
\end{pmatrix}.
\end{equation*}
Where $w_{\delta_i} = 2 \pi B ( \delta_i - \text{offset})$, and we have used a field of 400 MHz and an offset of 5 ppm.

Next, we need to calculate the contribution of the $J$-couplings to the Hamiltonian matrix. In this case, there is just one J-coupling:
$$
H_{J_{1,2}} = 2 \pi  J_{1,2} (S_x \otimes S_x + S_y \otimes S_y + S_z \otimes S_z).$$
Conversely:
$$H_{J_{1,2}} = 2 \pi  J_{1,2} \left[ \begin{pmatrix}
0 & \frac{1}{2} \\
\frac{1}{2} & 0
\end{pmatrix}
\otimes
\begin{pmatrix}
0 & \frac{1}{2} \\
\frac{1}{2} & 0
\end{pmatrix} + \begin{pmatrix}
0 & -\frac{i}{2} \\
\frac{i}{2} & 0
\end{pmatrix}
\otimes
\begin{pmatrix}
0 & -\frac{i}{2} \\
\frac{i}{2} & 0
\end{pmatrix} + \begin{pmatrix}
\frac{1}{2} & 0 \\
0 & -\frac{1}{2}
\end{pmatrix}
\otimes
\begin{pmatrix}
\frac{1}{2} & 0 \\
0 & -\frac{1}{2}
\end{pmatrix}\right].
$$

Summing up all the contributions to the Hamiltonian matrix, i.e.
\begin{equation*}
   H = H_{\delta_1} + H_{\delta_2} + H_{J_{1,2}} = w_{\delta_1}(S_z \otimes I_2 ) + w_{\delta_2}(I_2 \otimes S_z ) + 2 \pi J_{1,2}(S_x \otimes S_x + S_y \otimes S_y + S_z \otimes S_z),
\end{equation*}
we arrive at its final form:
\begin{equation*}
    H = \scalebox{1.0}{$ \begin{pmatrix}
1062.215 & 0 & 0 & 0 \\
0 & -4970.921 & 7.288 & 0 \\
0 & 7.288 & 4963.633 & 0 \\
0 & 0 & 0 & -1054.927 \\
    \end{pmatrix} $}
\end{equation*}

Now it is time to compute the eigenvalues and eigenvectors of that Hamiltonian matrix. We obtain them with quantum algorithms explained in the following sections without the need to construct the matrix. For now, just suppose that we have calculated them somehow. The eigenvalues are

\begin{equation*}
\begin{split}
\lambda_0 & = -4970.9263 \\
\lambda_1 & = -1054.927 \\
\lambda_2 & = 1062.215 \\
\lambda_3 & = 4963.6383
\end{split}
\end{equation*}

with the corresponding eigenvectors being

\begin{equation*}
\nu_0 = \begin{pmatrix}
0 \\
0.99999 \\
-0.00073 \\
0
\end{pmatrix}, \nu_1 = \begin{pmatrix}
0 \\
0 \\
0 \\
1
\end{pmatrix}, \nu_2 = \begin{pmatrix}
1 \\
0 \\
0 \\
0
\end{pmatrix} \text{ and } \nu_3 = \begin{pmatrix}
0 \\
-0.00073 \\
0.99999 \\
0
\end{pmatrix}.
\end{equation*}

The final part involves calculating the FID from the Hamiltonian matrix diagonalization (see, for example, \cite{hore2015nuclear}). First, we need to construct the diagonal form of the Hamiltonian matrix:

\begin{equation*}
D_H = \scalebox{1.0}{$ \begin{pmatrix}
-4970.9263 & 0 & 0 & 0 \\
0 & -1054.927 & 0 & 0 \\
0 & 0 & 1062.215 & 0 \\
0 & 0 & 0 & 4963.6383 \\
    \end{pmatrix} $},
\end{equation*}

and a matrix with the eigenvectors:

\begin{equation*}
V_H = \scalebox{1.0}{$ \begin{pmatrix}
0 & 0 & 1 & 0 \\
0.99999 & 0 & 0 & -0.00073 \\
-0.00073 & 0 & 0 & 0.99999 \\
0 & 1 & 0 & 0 \\
    \end{pmatrix} $}.
\end{equation*}

Then, if we want an FID with $d$ points of resolution, each point $\text{FID}(j)$ with $j \in \{0, \ldots, d-1 \}$ can be calculated as:
$$\text{FID}(j) = \text{Tr} \, (T(j) \, \sum_k S_{kx} ) + i \, \text{Tr} \, (T(j) \, \sum_k S_{ky} ),$$
where
$$T(j) = V_H e^{-i p_j^0 D} V_H^{\dagger} \left( \sum_k S_{kx} \right) V_H e^{i p_j^0 D} V_H^{\dagger},$$
$p_j^0 = j \cdot \text{SW}$, and SW is the spectral width desired for the resulting spectrum. After calculating the points of our FID and removing the unnecessary regions, we arrive at the result shown in Figure \ref{fig:enter-label}.

\begin{figure}[h!]
    \centering
    \includegraphics[scale=0.6]{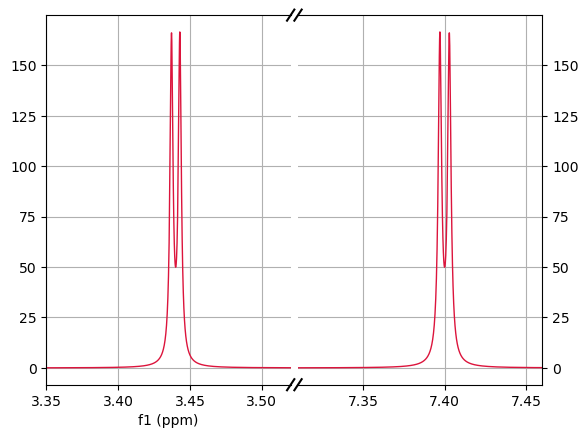}
    \caption{Classically calculated $^1$H-NMR spectrum of the sulfanol molecule}
    \label{fig:enter-label}
\end{figure}

Now that we have shown how to simulate the $^1$H-NMR spectrum of the sulfanol molecule, it is time to explain the process of obtaining the eigenvalues of its Hamiltonian matrix using a quantum computer with both the QPE and VQE algorithms.

The code for carrying simulations of the quantum algorithms was developed using the open-source toolkit Qiskit \cite{Qiskit}.

\subsection{Calculation of the Eigenvalues Using the QPE Algorithm}
In order to find the eigenvalues of the Hamiltonian of the sulfanol molecule, we have run a classical simulation of the QPE algorithm, using the Finisterrae III, the computer at CESGA in Santiago de Compostela, Spain. In this section, we proceed to enumerate our findings. The code used in the simulation can be found in \cite{rodriguez_coello_alexandre_2023_8337452}.

The workflow may be described as follows: we prepare the second register on a random initial state and we run the QPE circuit at most ten times. The first register contains 12 ancilla qubits, while in order to implement the $U$ gate we have used a Trotter number of 10 steps. For every outcome, we verify whether or not it is an eigenvalue by classically computing the  determinant of $H - \lambda_iI$. This step actually requires the construction of the matrix H, loosing the advantage of quantum computational approach of not needing to build the matrix. Additionally, a tolerance for the equality to zero must be selected. Every time an actual eigenvalue is successfully obtained, the second register collapse to its corresponding eigenstate. However, as noted in \cite{abrams1999quantum}, this information is trapped and it is not an easy task to access the full characterization of said state.

We first compute the scaled Hamiltonian choosing as our scale constant $C=24881.07$ (four times the upper bound for the greatest eigenvalue, as explained in Section \ref{ssec:qpe}). Then, we design the quantum circuit for the Trotterized $U$ gate using a number of 10 steps, obtaining an upper bound for the Trotter error of $\epsilon \approx 0.00033$.

For the scaled Hamiltonian of the sulfanol molecule, we have found three estimated phases with $t$ decimals as expected.
\begin{equation*}
\begin{split}
    \phi_1 & = -0.042480468750 \\
    \phi_2 & = 0.042724609375 \\
    \phi_3 & = 0.199462890625
    \end{split}
\end{equation*}
Since the sum of the eigenvalues of a matrix is equal to the trace of the matrix we may complete the set taking into account that $Tr(H) = 0$ (as it can be easily checked looking at its definition). 
\begin{equation*}
    \phi_0 = -0.199707031250
\end{equation*}
When compared to the classically computed eigenvalues, these estimations present a difference of less than $2^{-12}$, which agrees with the theoretical accuracy. 

By undoing the scale (see Section \ref{ssec:qpe}) we finally get the estimated eigenvalues of the Hamiltonian.
\begin{equation*}
    \begin{split}
        \lambda_0 & = -4968.925232949904 \\
        \lambda_1 & = -1056.959646128708 \\
        \lambda_2 & = 1063.0341268535858 \\
        \lambda_4 & = 4962.850752225026
    \end{split}
\end{equation*}

\subsection{Calculation of the Eigenvalues and Eigenstates Using the VQE}

Analogously to the previous section, here we present our findings when addressing the diagonalization of the sulfanol Hamiltonian by means of the VQE algorithm (in this case, the code can be found in \cite{rodriguez_coello_alexandre_2023_8337478}).

The XY-ansatz (see Section \ref{ssec:vqe}) will consist of only two parameters to be optimized. More precisely, the ansatz is represented by the following circuit:

\[
\Qcircuit @C= 1em @R= .7em {
&
\qw &
\gate{\parbox{1cm}{\centering $R_X$ \\ $-\pi /2$}} &
\qw &
\ctrl{1} &
\qw &
\qw &
\qw &
\ctrl{1} &
\gate{\parbox{1cm}{\centering $R_X$ \\ $\pi /2$}} &
\qw &
\gate{\parbox{1cm}{\centering $R_Y$ \\ $\pi /2$}} &
\qw &
\ctrl{1} &
\qw &
\qw &
\qw &
\ctrl{1} &
\gate{\parbox{1cm}{\centering $R_Y$ \\ $-\pi /2$}} &
\qw \\
&
\qw &
\gate{\parbox{1cm}{\centering $R_Y$ \\ $\pi /2$}} &
\qw &
\targ &
\qw &
\gate{\parbox{1cm}{\centering $R_Z$ \\ $2*\theta_{12}$}} &
\qw &
\targ &
\gate{\parbox{1cm}{\centering $R_Y$ \\ $-\pi /2$}} &
\qw &
\gate{\parbox{1cm}{\centering $R_X$ \\ $-\pi /2$}} &
\qw &
\targ &
\qw &
\gate{\parbox{1cm}{\centering $R_Z$ \\ $2*\theta_{21}$}} &
\qw &
\targ &
\gate{\parbox{1cm}{\centering $R_X$ \\ $\pi /2$}} &
\qw \\
}
\]

For every addend of the Hamiltonian, a quantum circuit is constructed consisting of this ansatz block plus a measure block.  This measured block would contain the necessary gates for changing the basis if needed. The circuit is then ran $1e4$ to estimate each expected value of the terms composing the total Hamiltonian.

We start from the same initial state as suggested in \cite{Wierichs2020avoidingminima}, which is:
\begin{equation*}
    \ket{\phi _0} = \frac{1}{\sqrt{2}}(0,1,-1,0)
\end{equation*}
Although in the original work this state is used for a Hamiltonian different from ours (a Heisenberg model not taking into account the individual terms due to the applied magnetic field) we show that even for our case a good result is still obtained. The parameters of the ansatz circuit must be initialized too and they are chosen to be both equal to zero. As for the optimizing process, we use the Constrained Optimization by Linear Approximation (COBYLA) method.

Finally, the result of the VQE calculation for the lowest eigenvalue is:
\begin{equation*}
    \lambda _0 = -4971.007.
\end{equation*}
Which is in good agreement with the classically calculated eigenvalue.

Since we also obtain the optimized parameters of the ansatz circuit, we may reconstruct the associated eigenvector. Using the formula for the XY-ansatz, we obtain the state:
\begin{equation*}
    v_0 = \begin{pmatrix}
    0 \\
    0.99999 \\
    0.00243 \\
    0
    \end{pmatrix}
\end{equation*}
Which is also similar to the classical result.

We attempt now to obtain higher eigenvalues via the folded spectrum method. As it was already explained, within this procedure one must compute the expected value of $(H -wI)^2$. Dealing with the square of a Hamiltonian involves a significant increase of Pauli terms. However, for the case of a 2 spin system, this growth does not appear. After performing the calculation for our specific Hamiltonian, it can be shown that we may decompose $(H -wI)^2$ in the exact same number of addends but with different coefficients:

\begin{equation*}
\begin{split}
    (H -wI)^2 &= (-2w w_{\delta_1} +2w_{\delta_2}J)Z \otimes I \\
    & + (-2w w_{\delta_2} +2w_{\delta_1}J)I \otimes Z \\
    &+ (-2J^2 -2wJ)X \otimes X \\
    &+ (-2J^2 -2wJ)Y \otimes Y \\
    &+ (2W_{\delta_1} w{\delta_2} -2wJ -2J^2)Z \otimes Z \\
    &+ (3J^2+ w_{\delta_1}^2 + w_{\delta_2}^2 +w^2)I \otimes I
\end{split}
\end{equation*}

Thereby, for this particular case, we may perform the same quantum procedure as for the first experiment and only modify the classical part where the total expected value is constructed. As for the $I  \otimes I$ term, the expected value of the identity operator with respect to a normalized state is trivially $1$. 

After successfully minimizing the cost function, the sum of $w$ and the square root of the result equals the eigenvalue closest to $w$.

Using the same initial state $\ket{\phi_0}$,  we perform the VQE algorithm for a certain number of $w$ values chosen based on the upper bound of  the eigenvalues of the Hamiltonian, aiming to find the largest eigenvalue. If $w$ is too low, the largest eigenvalue might not be the closest. On the other hand, a value for $w$ greater than the eigenvalue also results in a failed outcome of the algorithm. For appropriate $w$ values, we obtain good results with similar estimations for the largest eigenvalues.

Thus, thanks to the folded spectrum method we find:
\begin{equation*}
    \lambda_4 = 4965.511.
\end{equation*}
And for the eigenstate, using the optimized parameters, we obtain:
\begin{equation*}
    v_4 = \begin{pmatrix}
    0 \\
    0.02539 \\
    0.99968 \\
    0
    \end{pmatrix}.
\end{equation*}

For the intermediate eigenvalues, even for a good $w$ value, the result of the VQE algorithm is not satisfactory if the chosen initial state is not a good approximation of the ground state of the new $(H-wI)^2$. On the contrary, with good initial states and suitable $w$ values the intermediate eigenvalues are recovered. As for the negative eigenvalue, the folded spectrum method accepts $w<0$, so one can search for negative eigenvalues too. However, we found out that, in that case, the negative solution must be taken in the step of calculating the square root.

Then, the intermediate eigenvalues according to the VQE calculation are:
\begin{equation*}
    \lambda _3 = 1063.264  \hspace{3mm} v_3 = \begin{pmatrix}
        0.99999 \\
        0 \\
        0 \\
        -0.01027 \\
    \end{pmatrix}
\end{equation*} 
\begin{equation*}
     \lambda _2 = -1053.980  \hspace{3mm} v_2 = \begin{pmatrix}
        -0.0044 \\
        0 \\
        0 \\
        0.99999 \\
    \end{pmatrix}
\end{equation*}

As for the approach of modifying the cost function to take into account orthogonality of states, we were not able to obtain any good result for higher eigenvalues, which may imply that the XY-ansatz may be not suitable within this method for our particular case. When examining the matrix associated to the operation of the XY-ansatz over a state, it is clear to see that the ansatz cannot transform states of type $(a, 0, 0, b)$ into orthogonal states of type $(0,c,d,0)$, or viceversa, with this alternation being precisely present in the eigenstates of our Hamiltonian.

\subsection{Calculations on a Noisy Computer}

In this section we attempt to execute our model taking into account the noise present in real quantum computers. We have used the quantum computing services of CESGA, running the circuits using as a backend an emulator of their quantum computer, Qmio.

We introduce zero-noise extrapolation, or ZNE, as an available technique of error mitigation \cite{Giurgica_Tiron_2020}. Suppose we have built a quantum circuit to measure certain observable. In our case, we want to obtain the expectation value of our Hamiltonian in each iteration of the VQE algorithm. The current noise of the quantum hardware negatively affects the output obtained by the circuit. Assuming a certain proportion between the expectation value and the noise factor of the circuit, the solution proposed by ZNE consists on intentionally increased the noise level of the quantum circuit and then predict the value obtained in a noiseless situation by performing the fitting of an extrapolation model.

Thus, we run several circuits with intentionally increased noise level and then predict the output of a noiseless circuit by performing a regression taking into account the corresponding obtained values and the noise scale factors.

In order to scale the noise of the circuit we may increase the depth of the circuit via unitary folding. This method consists in adding $n$ pairs of a quantum gate and its adjoint. Since quantum gates are unitary operations, it holds:
$$U = (UU^{\dagger})^n$$
The logical behaviour of the circuit remains the same but the performance is expected to worsen since the number of physical operations have been scaled, and for each of these there exist fidelity errors. Also, one can increase the noise level with identity folding, where identity gates are added just to increase the time of execution and thus increasing the noise associated with depolarization.

In our method, we have applied unitary folding to each individual quantum gate composing the $XY$ ansatz (the other option was to fold the quantum circuit globally). For the rotation quantum gates used it is easy to see that $U(\theta)^{\dagger} = U(-\theta)$. As for the CNOT gate, the other quantum gate used to built the $XY$ ansatz, it is self adjointed.

We have run five quantum circuits with different scaled noise $\lambda$. At this point we clarify the quantification of the noise scale as the scale of the depth of the quantum circuit, i.e., how many times the number of physical operations have increased. For example, if for each quantum gate we add a new pair of gates, we have increased the overall size of the circuit by three times. Being $n$ the number of pairs we add for each quantum gate, the noise scale is $\lambda = 1 + 2n$. We supposed the noise increases polynomially with the size of the circuit and as for the extrapolation model we use Richardson extrapolation, using a polynomial fit of degree $m-1$ for our $m$ pairs of data. Other extrapolation models such as linear regression or exponential polynomial are of interested and should be studies in future works. We note that a drawback of Richardson extrapolation is the risk of overfitting the curve if the degree of the polynomial is too high, which is not the case in our work.

We have performed the calculations incorporating the ZNE correction to the expectation value calculated in each iteration of the VQE. The code implementing this technique can be found in \cite{rodriguezcoello_vqe_with_zne}.
\begin{figure} [h!]
    \centering
    \includegraphics[width=0.5\linewidth]{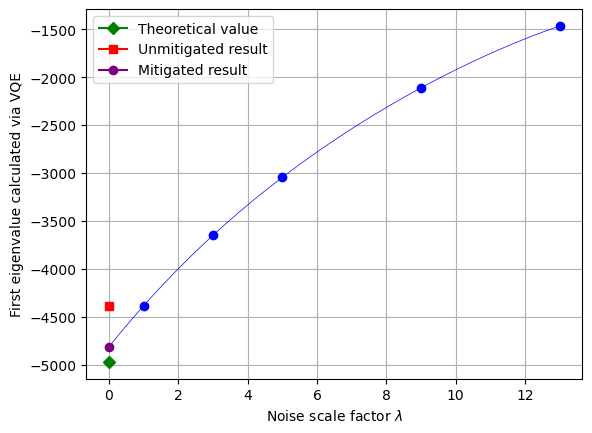}
    \caption{Comparison of the computed lowest eigenvalue with and without error mitigation. The extrapolation curve corresponds to the values of the last iteration of the optimization process.}
    \label{fig:enter-label}
\end{figure}

The theoretical value of the lowest eigenvalue was $\lambda_0 = -4970.921$ and the result obtained using the ideal noiseless simulator was $\lambda_0 = -4971.007$. When performing the VQE in the noisy simulator we initially obtain an unmitigated value of $\lambda_0 = -4384.837$. However, adding the zero noise extrapolation technique to the calculation of the expected value helps the VQE to return a much closer value of $\lambda_0 = -4813.889$, reducing the relative error from $11.79 \% $ to $3.16\%$ . 

Similarly, for the rest of eigenvalues, obtained applying the folded spectrum method to the VQE, the application of the ZNE improves the results. However, the relative error for the mitigated values is still too big and the effect of the noise is still considerable when comparing with the performance in the ideal simulator. A comparison can be seen in the table below.

\begin{table}[h!]
\centering
\begin{tabular}{l|l|l|l|l|}
\cline{2-5}
                                  & Theo. Value & Ideal Simulator & Unmitigated & Mitigated \\ \hline
\multicolumn{1}{|l|}{$\lambda_0$} & -4970.926   & -4971.007       & -4384.837   & -4813.889 \\ \hline
\multicolumn{1}{|l|}{$\lambda_1$} & -1054.927   & -1053.980       & -2313.336   & -1747.502 \\ \hline
\multicolumn{1}{|l|}{$\lambda_2$} & 1062.215    & 1063.264        & 2254.302    & 1670.105  \\ \hline
\multicolumn{1}{|l|}{$\lambda_3$} & 4963.638    & 4965.511        & 6002.497    & 5348.960  \\ \hline
\end{tabular}
\end{table}

Overall, the ZNE seems promising when reducing the noise and may be interesting to try other extrapolation model which may be more adjusted to the noise distribution of the quantum hardware. Specially in the context of future works with ansätze which require larger number of physical operations with the fidelity issues associated.

\subsection{Obtaining the Simulated NMR Spectrum}

\begin{figure}[h!]
    \centering
    \includegraphics[scale=0.6]{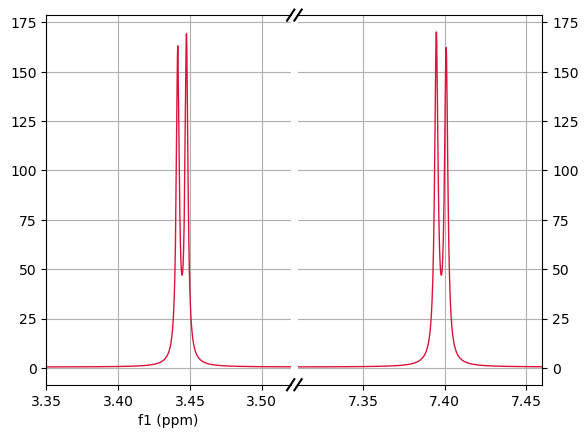}
    \caption{$^1$H-NMR spectrum of the sulfanol molecule using the $D_H$ and $V_H$ matrices calculated with our quantum algorithm}
    \label{fig:quant-sulf-spec}
\end{figure}

\begin{figure}[h!]
    \centering
    \includegraphics[scale=0.9]{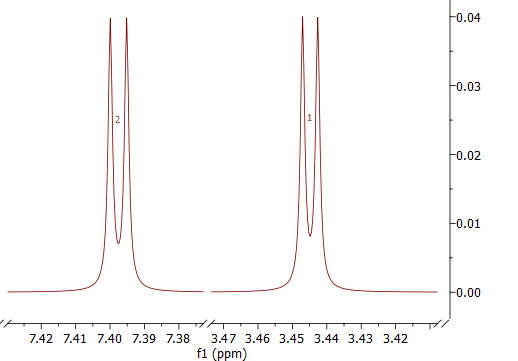}
    \caption{$^1$H-NMR spectrum of the sulfanol molecule simulated by MestReNova}
    \label{fig:sulfanol-simulated}
\end{figure}

Thanks to the VQE algorithm and the folded spectrum method, we have obtained the full set of eigenvalues of the Hamiltonian matrix of the sulfanol molecule, and we have also been capable of reconstructing the corresponding eigenvectors. This information has allowed us to build the $D_H$ and $V_H$ matrices and, after sending them to the program which calculates the FID \cite{sharman_gary_2023_8337288} and applying a Fourier transform, we have obtained the simulation of the $^1$H-NMR spectrum of sulfanol running a simulation of our quantum algorithm in a classical computer. The resulting simulation can be seen in Figure \ref{fig:quant-sulf-spec}, which is fairly close to the expected one (you can see the default simulation of the sulfanol $^1$H-NMR spectrum made by MestReNova in Figure \ref{fig:sulfanol-simulated}). 

\section{Conclusions and Future Research}

In this paper, we have shown how to simulate the $^1$H-NMR spectrum of a molecule's spin system, focusing on the diagonalization of its Hamiltonian matrix of said spin system. We have analyzed two alternative quantum algorithms, namely quantum phase estimation (QPE) and the variational quantum eigensolver (VQE) in order to improve this task.

For QPE, we have shown how the probability of obtaining a successful outcome depends on the initial state's overlap with  actual eigenstates. Thus, a protocol for selecting these initial states and deciding the number of attempts for each one is needed in order to optimize the QPE process. We have also raised the issue of efficiently verifying whether the obtained result is indeed an actual eigenvalue, as computing the determinant of the Hamiltonian matrix minus the identity matrix can be computationally expensive for high-dimensional matrices. Additionally, we have highlighted the challenge presented by degenerate eigenvalues, where QPE might yield similar estimations for the same eigenvalue, whether due to degeneracy or probabilistic reasons. We have suggested that access to information in the second register could potentially resolve this issue, a possibility which must be explored yet. It is again the difficulty or impossibility of characterizing the eigenstates stored in the second register the major drawback of QPE as a valid candidate for our goal.

The VQE offers a different approach to the diagonalization of the Hamiltonian matrix, with the benefits of eigenstates being obtainable and its applicability in current NISQ-era quantum computers, but with the inherent drawbacks of performing optimization processes. We have emphasized the critical role played by ansätze in the success of VQE while providing reference to various ansätze \cite{jattana2022assessment}, including the XY-ansatz \cite{you2022exploring}, which has been shown to be suitable even if the Hamiltonian includes individual terms due to an external magnetic field. Even though  VQE was originally designed to find the lowest eigenvalue and its corresponding eigenstate, it has the potential to discover additional eigenvalues and eigenstates through the folded spectrum method, which applies the VQE algorithm to a modified square Hamiltonian \cite{tazi2023folded}. Another extension consists in modifying the cost function, in which the goal shifts to minimizing the expectation value for state $k+1$ while ensuring that this state remains orthogonal to the previously determined states \cite{higgott2019vqe_all_states}. The former method may increase the measurement cost due to the requirement of computing the expected value of a squared Hamiltonian. Although for the simple case of  a system of two spins this is not an issue, it raises a concern for larger systems which should be addressed on further iterations of this project. Furthermore,  we have shown the importance of the initial state on which the parametric ansatz acts in order to obtain the correct result. As for the latter extension of modifying the cost function, no satisfactory results were obtained. This may have been caused due the incapacity of the XY-ansatz to arrive to a orthogonal state while starting from the ground state of our system. That being said, by combining the VQE algorithm and the folded spectrum method with suitable estimations of our Hamiltonian eigenstates, we were able to obtain the information needed to reconstruct the sulfanol molecule $^1$H-NMR spectrum with a good similarity with the classical one.

Future research should focus on further optimizing these algorithms and studying their viability for larger spin systems. Good estimation of ground states for the spin system Hamiltonian plays a major role in the success of the VQE. Transforming our Pauli encoded Hamiltonian into terms of Fermionic operators via the Jordan-Wigner transformation may be a promising path. Parallel to this suitable ground state estimation, the ansatz also influences the correct or incorrect result of the algorithm (a topic which is far from trivial), and more ansätze shall be reviewed and explored. Also, the optimization process must also be improved in order to tackle the mentioned issues of the local minima and the barren plateau problem.

In general, our quantum algorithms are formulated with ideal assumptions and some adjustments must be performed due to the current NISQ-era quantum processors. This is the case, for example, in the designing of an ansatz for the VQE algorithm where not only the specific characteristics of the Hamiltonian matrix to be diagonalized should be taken into account, but also the adaptation to the hardware. Specially in the context of the quantum computer available at the Galicia Supercomputing Center (CESGA), future research should focus in the adaptation of the quantum computing procedures in practical scenarios.

Additionally, other routes for the proposed problem may be studied beyond focusing on the process of the diagonalization the Hamiltonian, such as directly simulating the evolution of the spin system (see, for example, \cite{seetharam2021digital} and \cite{sels2020quantum}).
\section*{Acknowledgements}

The authors would like to thank the staff at CESGA for starting up this initiative, and especially Andrés Gómez, José Carlos Mouriño, Mariamo Mussa and Daniel Faílde, for their assistance during the various stages of this project and for their useful comments on an earlier draft of this manuscript. We would also like to thank Fran Pena from the University of Santiago de Compostela, for several clarifying discussions about quantum computing, and Gary Sharman from Mestrelab Research, for his help in the understanding of the whole process of simulating an NMR spectrum from a molecule's spin system. Finally, we would like to thank Carlos Cobas and Agustín Barba from Mestrelab Research, for supporting this project since its very inception.

This work was supported by the Axencia Galega de Innovación through the Grant Agreement ``Despregamento dunha infraestrutura baseada en tecnoloxías cuánticas da información que permita impulsar a I+D+i en Galicia'' within the program FEDER Galicia 2014-2020. Simulations on this work were performed using the Finisterrae III Supercomputer, funded by the project CESGA-01 FINISTERRAE III.

\bibliographystyle{siam}
\bibliography{references}

\end{document}